\magnification=1200
\mathsurround=2pt
\baselineskip=12pt plus 2pt

\def\cnt#1#2{\noindent\hangindent 4em\hangafter1\hbox to 4em{\hfil#1\quad}#2}

\bigskip

\bigskip

\bigskip

\bigskip

\centerline {\bf DYNAMICS AND SYMMETRIES}
\smallskip

\centerline {\bf OF A FIELD PARTITIONED BY}
\smallskip

\centerline {\bf AN ACCELERATED FRAME}
\bigskip

\bigskip

\centerline {Ulrich H. Gerlach}
\smallskip 
\centerline {Department of Mathematics}
\smallskip
\centerline {The Ohio State University}
\smallskip
\centerline {Columbus, Ohio  43210}
\bigskip

\bigskip

\bigskip

\centerline {\bf ABSTRACT}
\bigskip

The canonical evolution and symmetry generators are exhibited for a Klein-Gordon 
(K-G) system which has been partitioned by an accelerated coordinate frame into 
a pair of subsystems. This partitioning of the K-G system is conveyed to the 
canonical generators by the eigenfunction property of the Minkowski Bessel 
(M-B) modes.  In terms of the M-B degrees of freedom, which are unitarily 
related to those of the Minkowski plane waves, a near complete diagonalization 
of these generators can be realized.
\bigskip
\centerline {\bf TABLE OF CONTENTS}
\bigskip

\cnt {I.} {MOTIVATIONS}
\bigskip

\cnt {II.} {SYMMETRY GENRATORS FOR THE PARTITIONED SYSTEM}
\medskip

\cnt {~}{A.  Subsystem Generators:  Partial Diagonalization}
\smallskip

\cnt {~}{\quad \quad 1.  The Hamiltonian $H_I$}
\smallskip

\cnt {~}{\quad \quad 2.  Minkowski Bessel Modes}
\smallskip

\cnt {~}{\quad \quad 3.  The Hamiltonian $H_I$ (continued)}
\smallskip

\cnt {~}{\quad \quad 4.  The Transverse Subsystem Momenta $\vec P_I$ and 
$\vec P_{II}$.}
\medskip

\cnt{~}{B.  Total System Generators:  Complete Diagonalization}
\bigskip

\cnt {III.} {INERTIAL TRANSLATION GENERATORS FOR THE TOTAL SYSTEM}
\bigskip

\cnt {IV.} {CONCLUSION}	
\bigskip
\centerline {\bf I.  MOTIVATIONS}
\bigskip

\noindent Both within a classical and a quantum mechanical framework all 
inertial frames are equivalent.  This is expressed by the fact that the 
groundstates of a relativistic wave field in these frames are one and the 
same. Thus all inertial refrigerators cooling the wave field produce the same 
quantum state, the Minkowski vacuum.  
\medskip

1.  Considers however an accelerated refrigerator.
More specifically consider a pair 
of refrigerators accelerating linearly and uniformly into opposite directions. 
Their cooling operation results a quantum state that
is strikingly different from the Minkowski 
vacuum.  Its macroscopic manifestation can be characterized as a superfluid 
condensate of liquid light$^{1,2,3}$.
\medskip

These refrigerators are each situated in their own respective linearly 
uniformly accelerated 
coordinate frames.  The acceleration is such that these two frames are causally 
disjoint but jointly contain a Cauchy surface for all of Minskowski space-time.  
Consequently these two frames partition the wave field quantum system into two 
subsystems, which are respectively cooled by each of the two refrigerators.  
Each accelerated frame is static.  The time evolution of the wave field in a 
frame is governed by its Hamiltonian.  Any one of the field's stationary states 
is characterized by its conserved "boost" energy.  The ground state for the 
subsystem is the Fulling state$^4$.  There also is a Fulling state for the 
other subsystem.  The product of these two respective Fulling states is a 
quantum state for the total wave field system.  In the framework of condensed 
matter physics this total wave field state is called a condensed vacuum 
(= "Rindler"$^{5,6}$) state.  The motivation for this name is that on a macroscopic scale 
this state manifests itself as liquid light, a photon superfluid condensate$^1$.
\medskip

\noindent The framework within which condensed matter physics considerations 
arise is a Klein-Gordon system that has been partitioned by two accelerated 
frames into two subsystems whose symmetries and evolution bears a nonstandard 
relation to those of the total K-G system.  
\medskip

2.  One can obtain a general idea of the unusual relation between the pair of 
subsystems and the total system by contrasting the partioned K-G system, which 
has infinitely many quantized degrees of freedom, with an atomic system 
consisting of a pair of coupled angular momenta, which has only finitely many 
degrees of freedom.  For the finite atomic system the familiar Clebsch-Gordon 
technology establishes a unitary relation between the pair of subsystems and 
the total systems.  Guided by group representation theory, this technology allows 
one to classify and construct very systematically irreducible spaces of quantum states of the 
total system from the product of those of each subsystem.  For the atomic 
system the symmetry group is the rotation group.  It is unitarily implemented 
on the product states of the pair of subsystems as well as on those of the total 
system.  
\medskip

By contrast the symmetry group for the partitioned Klein-Gordon system is the 
two dimensional Poincare group (= "semidirect product" of translations and 
boosts on 2-D Lorentz space time$^7$).  It can be implemented on the dynamical variables of the total system.  It can not however 
be unitarily implemented on the quantum states of the system.  In 
particular translations in the Lorentz plane do not have a unitary 
representation:  the condensed vacuum state (= product of two Fulling vacua) is 
not translation invariant; it "breaks" the translational symmetry subgroup$^3$.  
A translation would produce what amounts to an excited state with an infinite 
number of quasiparticles.  It follows that this breaking of translational 
symmetry by the condensed vacuum state is an obstruction to constructing and  
classifying the K-G quantum states into irreducible unitary representation 
spaces of the two dimensional Poincare group in the Lorentz plane. 
\medskip

3.  To impose boundary conditions on a field, at some stage it 
invariably becomes necessary that one incorporate the space-time features of 
the reference frame.  Take, for example, a Klein Gordon field in an 
inertial reference frame.  This frame is equipped with the usual Minkowski 
coordinates and one therefore resolves the Klein-Gordon field into the familiar 
degrees of freedom, the Minkowski plane wave modes.  They express the 
translation invaraince of the frame, and they also make the implementation for 
the Feynman boundary conditions very easy. 
\medskip

If, by contrast, the field is described so as to incorporate the space-time features of a pair 
of accelerated frames, then one resolves the field into a different set of degrees 
of freedom, the Minkowski Bessel$^8$ (M-B) modes.
\medskip

These M-B modes readily leads to the total K-G system being partitioned 
into a pair of independent subsystems corresponding to the 
two accelerated frames.  Furthermore these modes, being Lorentz boost 
eigenfunctions, express the invariance of the union of the two accelerated 
frames under Lorentz boosts.  
\medskip

The incorporation of the space-time features of an accelerated frame can be 
extended from the  field to its canonical symmetry and evolution generators.  
These quadratic expressions are a vital ingredient for the quantum theory of the 
partitioned K-G system and as such give vital information about it.
\medskip

The purpose of this paper is to give for each of the two K-G subsystems the 
Minkowski Bessel mode expansion of their evolution and symmetry generators:  
 they are the Lorentz boost, i.e. moment of mass energy and the generators of translation, i.e. 
linear momentum, into the transverse directions (Section II.A).  The analogous 
generators are given for the total K-G system (Section II.B).  Similarly for 
the total K-G system we give such Minkowski Bessel mode expansions for all four 
translation generators (Section III)
\vfill\eject
\bigskip
\centerline {\bf II. SYMMETRY GENERATORS FOR THE PARTITIONED SYSTEM}.
\bigskip

If a wave field system is invariant under a space-time symmetry generated by a 
("Killing") vector field
$$\eqalignno {\xi^\mu {\partial \over {\partial x^\mu}}:\,\xi_{\mu ;\nu}\,&+
\,\xi_{\nu ;\mu}\,=\,0\,,&(2.1)\cr}$$
\medskip

\noindent then the associated conserved current satisfies
$$(\xi^\mu T^\nu_\mu);_\nu\,=\,0\,.$$
\medskip

\noindent Here $T^\nu_\mu$ is the stress-energy tensor which for a Klein-Gordon 
(K-G) system is
$$\eqalignno {T_{\mu \nu}\,&=\,\psi ,_\mu\,\psi ,_\nu\,-\,{1 \over {2}}g_{\mu 
\nu}(\psi ,_\alpha \psi ,\beta g^{\alpha \beta}\,+\,m^2\psi^2)\,&(2.2)\cr}$$
\medskip 

\noindent The conserved quantity  which remains invariant under the symmetry 
transformation is 
$$\eqalignno {\int\!\!\!\!\!\int\limits_{\scriptstyle space-like \atop \scriptstyle 
hypersurface }\!\!\!\!\!\int\,\xi^\mu \,&T^\nu_\mu\,d^3\Sigma_\nu\,.&(2.3)\cr}$$
\medskip

\noindent Consider the three Killing vector fields
$$\eqalignno {\xi^\mu_\tau\,x{\partial \over {\partial x^\mu}}\,&=\,{\partial \over 
{\partial t}}\,+\,t\,{\partial \over {\partial x}}\,\equiv\,{\partial \over 
{\partial \tau}}\cr
\xi^\mu_y\,{\partial \over {\partial x^\mu}}\,&=\,{\partial \over {\partial y}}\cr
\xi^\mu_z\,{\partial \over {\partial x^\mu}}\,&=\,{\partial \over {\partial z}}
&(2.4)\cr}$$
\medskip

\noindent and their corresponding canonical generators
$$\eqalignno {P_\tau\,&=\,\int\!\!\!\int\!\!\!\int\,\xi^\mu_\tau\,T^\nu_\mu\,d^3\Sigma_\nu\cr
P_y\,&=\,\int\!\!\!\int\!\!\!\int\,\xi^\mu_y\,T^\nu_\mu\,d^3\Sigma_\nu\cr
P_z\,&=\,\int\!\!\!\int\!\!\!\int\,\xi^\mu_z\,T^\nu_\mu\,d^3\,\Sigma_\nu &(2.5)\cr}$$
\medskip

\noindent where the integration is over a global space-like hypersurface.
\medskip

The Rindler coordinates $\xi$ and $\tau$ that mold themselves to these Lorentz 
and translation vector fields, Eqs. (2.4), are related to the MInkowski 
coordinates by 
$$\eqalign {t\,&=\,\pm\,\xi\,\sinh\,\tau\cr
x\,&=\,\pm\,\xi\,\cosh\,\tau\cr}$$
\medskip

\noindent (The coordinate $\xi$ is not to be confused with the vector symbol 
$\xi^\mu$).  Here $0<\xi <\infty , -\infty <\tau < \infty$ and the two signs 
refer to the two respective accelerated coordinate frames I and II.  In both of 
them the metric assumes the form
$$\eqalignno {ds^2\,&=\,-\xi^2\,dt^2\,+\,d\xi^2\,+\,dy^2\,+\,dz^2\,.&(2.6)\cr}$$
\medskip

\noindent The space-like integration surface of the canonical genrators, Eqs. 
(2.5), is partitioned by the pair of accelerated frames into the two parts
$$\lbrace \tau\,=\,const; -\infty >-\xi >0\rbrace~~{\rm which~lies~in}~~II$$  
\smallskip

and 
$$\lbrace \tau\,=\,const;\,0<\xi <\infty\rbrace~~{\rm which~lies~in}~~I\,.$$ 
\smallskip

All integration volume components $d^3\Sigma_\nu$ are therefore zero, except
$$\eqalignno {d^3\Sigma_\tau\,&=\,-\xi d\xi dy\,dz\,,&(2.7)\cr}$$ 
\medskip

\noindent which holds both in $II$ and in $I$.
\medskip

Inserting Eqs. (2.2), (2.4), and (2.7) into Eqs. (2.5) and keeping track of 
the orientation of the integration hypersurface in II, one obtains
$$\eqalignno {P_\tau\,&=\,H_{II}+H_I\cr
&={1 \over {2}}\int^0_\infty \int^\infty_{-\infty} \int^\infty_{-\infty}
\lbrace \cdots \rbrace \vert_{II}d\xi dy dz+{1 \over {2}}\int^\infty_0 
\int^\infty_{-\infty} \int^\infty_{-\infty}\lbrace \cdots \rbrace \vert_Id\xi dy dz
&(2.8)\cr}$$
\medskip

\noindent with
$$\lbrace \cdots \rbrace\,=\,\lbrace {1 \over {\xi}}(\psi,_\tau)^2+\xi 
\lbrack(\psi,_\xi)^2+(\psi,_y)^2+ (\psi,_z)^2+m^2\psi^2~\rbrack~ \rbrace\,,$$
\medskip

\noindent and 
$$\eqalignno {\vec P\,&=\,\vec P_{II}+\vec P_I\cr
&=\,\int^0_\infty\,\int^\infty_{-\infty}\,\int^\infty_{-\infty}\,\lbrace 
{1 \over {\xi}}\psi ,_\tau\,\vec \nabla\,\psi \rbrace d\xi dy dz+
\int^\infty_0\,\int^\infty_{-\infty}\,\int^\infty_{-\infty}\lbrace 
{1 \over {\xi}}\,\psi ,_\tau\,\vec \nabla\,\psi \rbrace d\xi dy dz
&(2.9)\cr}$$
\medskip

\noindent with 
$$\vec P \equiv (P_y, P_z),~~\vec \nabla \psi \equiv ({\partial \psi \over 
{\partial y}}, {\partial \psi \over {\partial z}})\,.$$
\medskip

\noindent Thus the Lorentz boost generator and the transverse translation 
generator decompose into pats which affect the pair of K-G subsystems
\medskip

\noindent with
$$\vec P\equiv (P_y, P_z),~~ \vec \nabla \psi \equiv ({\partial \psi \over 
{\partial y}}, {\partial \psi \over {\partial z}})\,.$$
\medskip

\noindent The physical interpretation of the boost generator $P_\tau$ depends 
on the frame.  Relative to the accelerated frame  $I$, $H_I$ is the conserved 
energy because $\tau$ is the time parameter.  Similarly $H_{II}$ is the 
conserved energy for the accelerated frame $II$.  Relative to the inertial 
frame the interpretation is different.  The Minkowski frame integral, 
obtained from Eqs. (2.4) and (2.5),
$$P_\tau\,=\,\int\!\!\!\int\limits_{t=0}\!\!\!\int\,x\,T_{00}dx dy dz$$
\medskip

\noindent implies that $P_\tau$ is the total moment of mass-energy of the 
K-G system around the event $t=x=0$.  By contrast $H_I$ is the partial moment 
of mass energy of subsystem $I$.
\medskip

\noindent  A. {\it Substystem Generators:  Partial Diagonalization}
\medskip
We shall reexpress the subsystem I evolution generator
$$\eqalignno {H_I\,&=\,{1 \over {2}}\int^\infty_0\,\int^\infty_{-\infty}
\int^\infty_{-\infty}\lbrace {1 \over {\xi}}(\psi ,_\tau )^2+\xi \lbrack~ 
\psi ,_\xi )^2+(\psi ,_y)^2+(\psi ,_z)^2+m^2\psi^2\rbrack ~~\rbrace ~d\xi dy dz~
&(2.10)\cr}$$
\medskip

\noindent and the two subsystem $I$ momentum components 
$$\eqalignno {\vec P_I\,&=\,\int^\infty_0\,\int^\infty_{-\infty}\,
\int^\infty_{-\infty}\lbrace ~{1 \over {\xi}}\psi,_\tau \vec \nabla \psi~
\rbrace ~d\xi dy dz\,,&(2.11)\cr}$$
\medskip

\noindent in terms of the Minkowski Bessel modes.    
\medskip

\noindent 1.)  $\underline {\rm The\,\,Hamiltonian\,\,H_I}$
\medskip
 
First focus on the observer Hamiltonian $H_I$, Eq. (2.10).  The task is 
simplified considerably with the help of the of the K-G field equation.  
It can be used to eliminate reference to derivatives with respect to the 
non-cyclic coordinate $\xi$:
$$\xi (\psi ,_\xi)^2\,=\,{\partial \over {\partial \xi}}(\xi \psi \psi ,_\xi)\,-\,
{1 \over {\xi}}\psi \psi,_{\tau \tau}\,+\,\xi \psi ~\lbrack \psi ,_{yy}\,+\,
\psi ,_{zz}\,-\,m^2\psi \rbrack \,.$$
\medskip

\noindent Thus the subsystem Hamiltonian $H_I$,  reduces to
$$\eqalignno {H_I\,&=\,{1 \over {2}}\int^\infty_0
\int^\infty_{-\infty}\int^\infty_{-\infty}\lbrace ~\,(\psi ,_\tau \psi  ,_\tau\,
-\,\psi \psi ,_{\tau \tau})/\xi\,+\,(\psi ,_y\psi ,_y\,+\,\psi \psi ,_{yy})
\xi \cr
\cr
&+\,(\psi ,_z\psi ,_z\,+\,\psi \psi ,_{zz})\xi ~\rbrace ~d\xi dydz\,.
&(2.12)\cr}$$
\medskip

\noindent Here we have performed an integration by parts (w.r.t. $\xi$) and 
dropped the surface terms.  Now introduce the K-G field expanded in terms of 
the globally defined Minkowski-Bessel modes \hfil\break $B^{\pm}_\omega (kU,kV)$.  
\medskip

\noindent 2.  $\underline {\rm The\,\,Minkowski\,\,Bessel\,\,Modes}$ 
\medskip

These mode are defined on two dimensional Lorentz space-time spanned by 
$t={1 \over {2}}(V+U)$ and $ x= {1 \over {2}}(V-U)$ and are given by $^8$
$$\eqalignno {B^{\pm}_\omega (kU,kV)\,&=\,{1 \over {2\pi }}\int^\infty_{-\infty}
e^{\mp i(\omega_kt-k_xx)}\,\,e^{-i\omega \theta} d\theta\cr
&={1 \over {2\pi}}\int^\infty_{-\infty}\int^\infty_{-\infty}\,exp\,\lbrack 
\mp ik(Ue^{\theta}+Ve^{-\theta})/2\rbrack~~ e^{-i\omega \theta}\,d\theta 
&(2.13)\cr}$$
\medskip

\noindent with
$$\eqalignno {\omega_k\,&=\,k\,\cosh\,\theta\cr
k_x\,&=\,k\,\sinh\,\theta\cr
k\,&=\,\sqrt {k^2_y+k^2_z+m^2}\,>\,0&(2.14)\cr}$$
\medskip

\noindent Their coordinate representatives for $I$ and $II$ are
$$\eqalignno {B^{\pm}_\omega (kU,kV)\,&=\,{1 \over {\pi}} K_{i\omega }(k\xi )
e^{-i\omega \tau}\,\left\{ \matrix {e^{\pm \pi \omega /2} &in &I\cr
                                    e^{\mp \pi \omega /2} &in &II\cr}\right.\,,
&(2.15)\cr}$$
\medskip

\noindent and they have the following relevant properties:
\medskip

\quad (i)
$$\eqalignno {{\partial \over {\partial \tau}}B^{\pm}_\omega \,&=\,
-\,i\omega B^{\pm}_\omega&(2.16a)\cr}$$
\medskip

\quad (ii) 
$$\eqalignno {(B^+_\omega )^\ast\,&=\,B^-_{-\omega} &(2.16b)\cr}$$
\medskip

\quad (iii)
$$\eqalignno {\int^\infty_0\,B^+_\omega\,B^{\pm}_{\pm \omega '}\,
{d\xi \over {\xi}}\,&=\,{e^{-i(\omega \pm \omega ')\tau} \over {2\omega 
\sinh \pi \omega }}\lbrack \delta (\omega + \omega ')+\delta (\omega - \omega ')
\rbrack \left \{ \matrix {e^{\pi (\omega + \omega ')/2}&in&I\cr
                          e^{-\pi (\omega +\omega ')/2}&in&II\cr}\right \}
&(2.16c)\cr}$$
\medskip

\noindent The last integral follows from 
$$\int^\infty_0\,K_{i\omega}(k\xi )\,K_{i\omega '}(k\xi ) {d\xi \over {\xi}}\,
=\,{\pi^2 \over {2\omega \sinh \pi \omega}}\lbrack \delta (\omega + \omega ')+\delta (\omega _- \omega ')\rbrack$$
\medskip

\noindent and Eq. (2.15)
\medskip

\noindent  3.  $\underline {\rm The\,\,Hamiltonian\,\,H_I\,\,(continued).}$
\medskip

Expanded in terms of these modes the K-G field is 
$$\eqalignno {\psi\,&=\,(1/\sqrt {2})\int^\infty_{-\infty}
\int^\infty_{-\infty}\int^\infty_{-\infty}a_{\omega k}B^+_\omega 
{e^{i(k_yy+k_zz)} \over {2\pi}}\,+\,a^+_{\omega k}~(B^+_\omega )^\ast\,
{e^{-i(k_yy+k_zz)} \over {2\pi }}\,d \omega dk_y\,dk_z&(2.17)\cr}$$
\medskip

\noindent The last two integrals in Eq. (2.12) (involving the spatial 
derivations) vanish.  This is because with $\psi (x)$ given by Eq. (2.17), the 
first one, say, is proportional to
$$\int^\infty_{-\infty}\int^\infty_{-\infty}\int^\infty_{-\infty}\lbrack 
\pm k'_yk_y\,+\,k'_yk'_y\rbrack~ \delta (k_y\pm k'_y)~dk_ydk'_y\,=\,0.$$
\medskip

The second one involving $k_z$ and $k'_z$ also vanishes.  Thus the observer
Hamiltonian $H_I$ reduces to
$$\eqalignno {H_I\,&=\,{1 \over {2}}\int^\infty_0
\int^\infty_{-\infty}\int^\infty_{-\infty}{1 \over {2}}(\psi ,_\tau 
\psi ,_\tau - \psi \psi  ,_{\tau \tau})\xi^{-1}\,d\xi dydz\,.&(2.18)\cr}$$
\medskip

\noindent This Hamiltonian can be readily expressed in terms of the quantum 
operators $a_{\omega k}$ and $a^+_{\omega k}$.  The field $\psi$, Eq. (2.17), 
is given by
$$\psi \,=\,\psi^+\,+\,\psi^-\,=\,\psi^+\,+\,c.c.$$
\medskip

\noindent where
$$\psi^+\,=\,(1/\sqrt {2})\int^\infty_{-\infty}
\int^\infty_{-\infty}\int^\infty_{-\infty}\,B^+_\omega (kU,kV)~ a_{\omega k}\,
{e^{ik_yy+ik_zz} \over {2\pi}}d\omega dk_ydk_z$$
\medskip

Thus the Hamiltonian $H_I$, Eq. (2.18), becomes 
$$\eqalignno {H_I\,&=\,{1 \over {2}}\int^\infty_0
\int^\infty_{-\infty}\int^\infty_{-\infty}\lbrace ~\psi^+_{,\tau}
(\psi ,^+_\tau \,+\,\psi ,^-_\tau )\,-\,
\psi ,^+_{\tau \tau}(\psi^+\,+\,\psi^-)\,+\,c.c.\rbrace \xi^{-1}~d\xi dydz\cr
\cr
&=\,{1 \over {2}}\int^\infty_0\xi^{-1}d\xi \int^\infty_{-\infty}
\int^\infty_{-\infty}\int^\infty_{-\infty}d\omega d^2k\int^\infty_{-\infty}
\int^\infty_{-\infty}\int^\infty_{-\infty}d\omega 'd^2k'\cr
\cr
&{1 \over {2}}\lbrace~ (-)B^+_\omega ~\omega \,a_{\omega k}\,\omega '~\lbrack 
B^+_{\omega '}\,a_{\omega 'k'}\,\delta^2(k+k')+(B^+_{\omega '})^\ast\, 
a^+_{\omega 'k'}\,\delta^2(k-k' )\rbrack\cr
\cr
&-\,(-)B^+_\omega~ \omega^2 a_{\omega k}~\lbrack B^+_{\omega '}\,a_{\omega 'k'}\,
\delta^2(k+k')+(B^+_{\omega '})^\ast \,a^+_{\omega 'k'}\,\delta^2(k-k')\rbrack 
+c.c.~\rbrace &(2.19)\cr\cr}$$
\medskip

\noindent where we used the eigenfunction property, Eq. (2.16a)
\medskip

The Minkowski Bessel modes $B^{\pm}_\omega$ are even functions of $k_y$ and 
$k_z$.  This follows from  Eqs. (2.13) and (2.14).  Thus there is no 
difficulty in doing the integration $\int \int \cdots d^2k'$.   The step next 
to the last is to perform the integration over the remaining spatial coordinate 
$\xi$.  Using Eqs. (2.16b) together with (2.16c) one obtains the Klein-Gordon 
subsystem Hamiltonian $H_I$, 
$$\eqalignno {H_I\,&=\,{1 \over {2}} \int^\infty_{-\infty}\int^\infty_{-\infty}
\int^\infty_{-\infty} d\omega d^2k\int^\infty_{-\infty}d\omega '{1 \over 
{2\omega \sinh \pi \omega }}\cr
\cr
&{1 \over {2}}\lbrace~ (-\omega \omega '+\omega^2)e^{-i(\omega +\omega ')\tau}~~
e^{\pi (\omega +\omega ')/2}\lbrack ~~\delta (\omega -\omega ')+\delta(\omega +
\omega ')\rbrack ~a_{\omega k}a_{\omega '-k}\cr
\cr
&+(\omega \omega '+\omega^2)e^{-i(\omega -\omega ')\tau}~~e^{\pi (\omega +\omega ')
/2}\lbrack~~ \delta (\omega - \omega ')+\delta (\omega +\omega ')\rbrack ~a_
{\omega k}a^+_{\omega ' k}+c.c.\rbrace \,,\cr
\cr
&=\,{1 \over {2}}\int^\infty_{-\infty}\int^\infty_{-\infty}\int^\infty_{-\infty}
d\omega d^2\,k{\omega \over {2\omega \sinh\,\pi \omega}}\lbrace ~a_{\omega k}~
a_{-\omega -k}+a^+_{\omega k}~a^+_{-\omega -k}\cr
\cr
&+\,e^{\pi \omega }(a_{\omega k}a^+_{\omega k}+a^+_{\omega k}a_{\omega k})
\rbrace\,.&(2.20)\cr}$$
\medskip

\noindent Without much ado one can readily see that the expansion for $H_{II}$, 
Eq. (2.8), is nearly the same.  Equations (2.15) and (2.16c) dictate the replacement
$$e^{\pi \omega}\,\to\,e^{-\pi \omega}$$
\medskip

\noindent Thus $H_{II}$ decomposes into an analogous mode sum.
$$\eqalignno {H_{II}\,=\,-{1 \over {2}}\int^\infty_{-\infty}\int^\infty_{-\infty}
\int^\infty_{-\infty}d\omega d^2&k{\omega \over {2\sinh\pi \omega}}\lbrace 
~a_{\omega k}a_{-\omega -k}+a^+_{\omega k}a^+_{-\omega    -k}\cr
\cr
&+\,e^{-\pi \omega}(a_{\omega k}a^+_{\omega k}+a^+_{\omega k}a_{\omega k})
\rbrace \,.&(2.21)\cr}$$
\medskip

\noindent 4.  $\underline {\rm The\,\,Transverse\,\,Subsystem\,\,Momenta\,\,
\vec P_I\,\,and\,\,\vec P_{II}}$.
\medskip

To express $\vec P_I$, Eq. (2.11), in terms of the Minkowski Bessel degrees of 
freedom in Eq. (2.17) is trivial compared to $H_I$.  The result is, nevertheless,
just as interesting.  One obtains 
$$\eqalignno {\vec P_I\,=\,{1 \over {2}}\int^\infty_{-\infty}
\int^\infty_{-\infty}\int^\infty_{-\infty}d\omega d^2 &k{\vec k \over {2\sinh
\pi \omega}}\lbrace ~a_{\omega k}~a_{-\omega -k}+a^+_{\omega k}~a^+_{-\omega -k}\cr
\cr
&+\,e^{\pi \omega }(a_{\omega k}~a^+_{\omega k}+a^+_{\omega k}~a_{\omega k})
\rbrace\,.&(2.22)\cr}$$
\medskip

\noindent Similarly for $\vec P_{II}$ in Eq. (2.9) one obtains
$$\eqalignno {\vec P_{II}\,=\,-{1 \over {2}}\int^\infty_{-\infty}
\int^\infty_{-\infty}\int^\infty_{-\infty}d\omega d^2&k{\vec k \over {2\sinh
\pi \omega}}\lbrace~a_{\omega k}a_{-\omega -k}+a^+_{\omega k}a^+_{-\omega -k}\cr
\cr
&+\,e^{-\pi \omega }(a_{\omega k}a^+_{\omega k}+a^+_{\omega k}a_{\omega k})
\rbrace\,.&(2.23)\cr}$$
\medskip

\noindent The K-G subsystem I is characterized by three commuting conserved canonical 
generators.  As one can see from Eqs. (2.20) and (2.22) the Minkowski Bessel 
modes bring about a nearly complete but not quite total simultaneous 
diagonalization.
\medskip

The analogous result for K-G subsystem II is given by Eqs. (2.21) and (2.23).
\medskip

\noindent B. {\it Total System Generators:  Complete Diagonalization.} 
\medskip

The generator of boosts for the total Klein-gordon system around $x=t=0$ is 
merely the sum of the generatos, Eqs. (2.20) and (2.21), 
$$\eqalignno {P_\tau\,&=\,H_I+H_{II}\cr
&+{1 \over {2}}\int^\infty_{-\infty}\int^\infty_{-\infty}\int^\infty_{-\infty}\,
\omega (a_{\omega k}a^+_{\omega k}+a^+_{\omega k}a_{\omega k})d\omega 
dk_ydk_z&(2.24)\cr}$$
\medskip

\noindent Similarly the generator of transverse translations for the total K-G 
system is the sum of Eqs. (2.22) and (2.23),
$$\eqalignno {\vec P\,&=\,\vec P_I\,+\,\vec P_{II}\cr
&=\,{1 \over {2}}\int^\infty_{-\infty}\int^\infty_{-\infty}\int^\infty_{-\infty}
\,\vec k\,(a_{\omega k}a^+_{\omega k}\,+\,a^+_{\omega k}a_{\omega k})d\omega 
dk_y dk_z&(2.25)\cr}$$
\medskip

\noindent Thus the Minkowski-Bessel (boost eigenfunction) degrees of freedom 
bring about a complete diagonalization fo the boost generator and the 
transverse momentum of the total K-G System.

\bigskip

\centerline {\bf III.  TRANSLATION GENERATORS FOR THE TOTAL SYSTEM}.
\medskip

An accelerated coordinate frame partitions the total Klein Gordon system into a 
pair of subsystems.  This partitioning is incorporated into the very nature of 
the Minkowski Bessel modes (by virtue of the event horizons $U=0$ and $V=0$ being 
the coordinate chart boundaries).  Nevertheless these modes are still unitarily 
related to the Minkowski plane wave modes (see Eq. (2.7).  (Physically this 
means that emission and /or absorption matrix elements computed relative to the 
plane wave particle quantum state basis are equal to those computed relative 
to the quantum state basis associated with Minkowski Bessel mode particles.)  
Symmetry operations and dynamical evolution (in classical variables or quantum 
states) bring about changes which are generated by the canonical generators.  
Suppose these generators are expressed in terms of the Minkowski-Bessel degrees 
of freeedom.  Then the acceleration induced partitioning of the total 
K-G system can manifest itself classically and quantum mechanically through 
these canonical generators.
\medskip

Then canonical generator (for the total K-G system) which are of interest 
are the total inertial energy and momentum, the translation generators,
$$\eqalignno {P_t\,&=\,\int \int \int \xi^\mu_t\,T^0_\mu\,d^3x\cr
P_x\,&=\,\int \int \int \xi^\mu_x\,T^0_\mu\,d^3x\cr
P_y\,&=\,\int \int \int \xi^\mu_y\,T^0_\mu\,d^3x\cr
P_z\,&=\,\int \int \int \xi^\mu_z\,T^0_\mu\,d^3x&(3.1)\cr}$$ 
\medskip

\noindent and the total moment of mass energy, the Lorentz boost generator,
$$\eqalignno {P_\tau\,&=\,\int \int \int \xi^\mu_\tau\,T^\nu_\mu\,d^3 
\Sigma_\nu &(3.2)\cr}$$
\medskip

The purpose of this subsection is to expand the inertial energy and momentum in 
terms of the Minkowski Bessel modes.  This goal we can achieve in three steps.
\medskip

\quad (i)  Notice that all these generators can be obtained from a single scalar 
as follows:  
\medskip

Recall the familiar plane we expansion of each generator
$$P_\mu\,=\,{1 \over {2}} \int \int \int k_\mu (a_{\vec k}a^+_{\vec k}+
a^+_{\vec k}\,a_{\vec k})d^3k\quad\quad\quad\quad\mu = 0,1,2,3$$
\medskip
\noindent where $(k_0, k_1, k_2, k_3)=(-\omega _k, k_x, k_y, k_z )$
\medskip

\noindent and $\omega_k\,=\,\sqrt {k^2_x+k^2_y+k^2_z+m^2}\,>\,0$
\medskip

Consider the "generating" scalar
$$\eqalignno {{\cal {P}}(x)\,&=\,{1 \over {2}}\int \int\int e^{ik_\mu x^\mu}
(a_{\vec k}\,a^+_{\vec k}+a^+_{\vec k}\,a_{\vec k})d^3k\,.&(3.3)\cr}$$
\medskip

\noindent It follows that
$$\eqalignno {P_\mu\,&=\,{1 \over {i}}\,{\partial \over {\partial x^\mu}}
{\cal {P}}(x)\vert_{x=0}~~~~\mu =0,1,2,3 &(3.4)\cr}$$
\medskip

\noindent are precisely the four translation generators, 
$$(P_t, P_x, P_y, P_z)\,=\,(-P_0, P_1, P_2, P_3)\,.$$
\medskip

\noindent Furthermore
$$\eqalignno {P_\tau\,&=\,-\lbrack\,x{\partial \over {\partial t}}\,+\,t\,
{\partial \over {\partial x}}\,\rbrack {\cal {P}}(x^\nu )\vert_{x^\nu =0}\cr
&=\,-\,{\partial \over {\partial \tau }}{\cal {P}}(x^\nu )\vert_{x^\nu = 0}
&(3.5)\cr}$$
\medskip

\noindent Thus all four generators can be obtained by taking appropriate 
derivatives of the generating scalar ${\cal {P}}(x)$.
\medskip

\quad (ii)  Relabel the modes in terms of the positive mass shell parameters 
$\theta ,k_y, k_z$, which are defined by
$$\eqalignno {\omega_k\,&=\,k\,\cosh\,\theta ; k_x\,=\,k\,\sinh\,\theta\cr
k\,&=\,\sqrt {k^2_y+k^2_z+m^2}&(3.5)\cr}$$
\medskip

\noindent and in terms of which the boost (in the $x-t$ plane) invariant measure is 
$$\eqalignno {{d^3k \over {\omega_k}}\,&=\,d\theta dk_y\,dk_z&(3.6)\cr}$$
\medskip

\noindent In terms of these parameters the Klein-Gordon Minkowski plane wave mode 
expansion is 
$$\eqalignno {\psi (x)\,&=\,\int^\infty_{-\infty}\int^\infty_{-\infty}
\int^\infty_{-\infty}\,(a_{\theta k}\,f_{\theta k}+a^+_{\theta k}
f^\ast_{\theta k})d\theta dk_y dk_z,&(3.7)\cr}$$
\medskip

\noindent where
$$\eqalignno {f_{\theta k}\&=\,\lbrack 2(2\pi )^3\rbrack^{-1/2}\,
e^{-i(\omega_kt-k_xx)}\,
e^{i(k_yy+k_zz)}\cr
&=\,\lbrack 2(2\pi )^3\rbrack^{-1/2}\,e^{-ik\lbrack (t-x)e^\theta +(t+x)
e^{-\theta}\rbrack/2}\,e^{i(k_yy+k_zz)}&(3.8)\cr}$$
\medskip

The familiar dynamical plane wave amplitude $a_{\vec k}$ is related to 
the reparametrized one by
$$\eqalignno {a_{\theta k}\,&=\,\sqrt {k\cosh\,\theta}\,a_{\vec k}&(3.9)\cr}$$
\medskip

\noindent Insert Eqs. (3.6), (3.8) and (3.9) into (3.3), and the generating 
scalar becomes
$$\eqalignno {{\cal {P}}(x)\,&=\,{1 \over {2}}\int^\infty_{-\infty}\int^\infty_{-\infty}
\int^\infty_{-\infty}\,exp\lbrack - ik(Ue^\theta +Ve^{-\theta})/2\rbrack 
e^{i(k_yy+k_zz)}\,(a_{\theta k}a^+_{\theta k}+a^+_{\theta k}a_{\theta k})\,
d\theta dk_ydk_z\cr &&(3.10)\cr}$$
\medskip

\noindent where $U=t-x;~~V=t+x$.
\medskip

\quad (iii)  Implement the unitary transformation $e^{-i\omega \theta}/\sqrt 
{2\pi}$.  This yields the Minkowski Bessel modes Eq. (2.13) from the relabelled 
plane wave modes in Eq. (3.8).  It also gives the Minkowski Bessel dynamical 
amplitude
$$ a_{\omega k}\,=\,\sqrt {1/2\pi}\int^\infty_{-\infty}a_{\theta k}\,e^{i\omega \theta}d\theta$$
\medskip

\noindent In terms of these amplitudes and these M-B modes the generating scalar is
$$\eqalignno {{\cal {P}}(x)\,&=\,{1 \over {2}}\int^\infty_{-\infty}\int^\infty_{-\infty}
\int^\infty_{-\infty}\int^\infty_{-\infty}\,B^+_{\omega - \omega '}(kU,kV)\,
e^{i(k_yy+k_zz)}(a_{\omega k}a^+_{\omega 'k}\,+\,a^+_{\omega k}\,a_{\omega 'k})
d\omega d\omega ' dk_ydk_z\cr &&(3.11)\cr}$$
\medskip

\noindent From this generating scalar one can obtain the total K-G energy and
momentum expressed in terms of the M-B degrees of freedom.  One must use 
Eq. (3.4) to do this.
\medskip

One readily recovers the total transverse momentum, Eq. (2.25), from the 
generating scalar, Eq. (3.11).  Using Eq. (3.4) together with
$$B^{\pm}_{\omega - \omega '}(0,0)\,=\,\delta (\omega - \omega '),$$
\medskip

\noindent one obtains
$$\eqalign {\vec P\,&=\,\vec \nabla \,{\cal {P}}(x^\nu)
\vert_{x^\nu =0}\cr
&=\,{1 \over {2}}\int^\infty_{-\infty}\int^\infty_{-\infty}\int^\infty_{-\infty}\,
\vec k\,(a_{\omega k}a^+_{\omega k}+a^+_{\omega k}\,a_{\omega k})~~d\omega dk_ydk_z\cr
&=\,\vec P_I+\vec P_{II}\cr}$$
\medskip

\noindent  which agrees with Eq. (2.25).  
\medskip

Similarly one recovers the total moment of mass energy of the field, Eq. (2.24), 
by inserting (2.16a) into (3.5),
$$\eqalign {P_\tau\,&=\,-\,{\partial \over {\partial \tau}}\,{\cal {P}}
(x^\nu)\vert_{x^\nu =0}\cr
&=\,{1 \over {2}}\,\int^\infty_{-\infty}\int^\infty_{-\infty}
\int^\infty_{-\infty}\,\omega\,(a_{\omega k}a^+_{\omega k}\,+\,
a^+_{\omega k}\,a_{\omega k})d\omega dk_ydk_z\cr}$$
\medskip

One can also exhibit the expansions of the longitudital momentum components 
$P_t$ and $P_x$ in terms of the Minkowski-Bessel degrees of freedom, but we 
shall delay doing this until we use them to discuss the longitudinal 
translational symmetry breaking of the K-G system.
\bigskip
\centerline  {\bf IV.  CONCLUSION}
\medskip

The basic result of this paper is that the total Klein-Gordon system,
with its infinite number of dynamical degrees of freedom, is
partitioned by an (linearly uniformly) accelerated coordinate frame
into a pair of independent dynamical subsystems.  This result has been
made precise by exhibiting the canonical evolution and symmetry
generators for these subsystems (as well as for the total system) in
terms of the Minkowski-Bessel degrees of freedom.  They are those
consitutents of the K-G system which exhibit its partitioning but are
still unitarily related to the plane wave degrees of freedom.  This
Minkowski Bessel mode decomposition of the canonical generators is at
the basis of the quantum dynamics of the acceleration partioned K-G
system.  These dynamics are discussed elsewhere $^2$.
\bigskip
\newcount\bibno\bibno=1
\def\bib#1\par{\hangindent 3em\hangafter 1\noindent\hbox to 3em{\hfil\number
\bibno.\quad}#1\par\smallskip\advance\bibno by 1}

\centerline {\bf References}
\medskip

\bib {U. H. Gerlach, "The Rindler Condensate:  Ground State in an Accelerated 
Frame"  \hfill\break in $\underline {\rm Proceedings\,\,of\,\,4th\,\,Marcel\,\,Grossmann\,\,
Meeting\,\,on\,\,General\,\,Relativity}$,\hfill\break R. Ruffini (ed.)  (Elsevier Science 
Publishers B.V., 1986) P1129-1138.}
\smallskip

\bib {U. H. Gerlach, "Liquid Light and Accelerated Frames'', Annales de 
l' I.H.P {\bf 49}, 397-401 (1988).}
\smallskip

\bib {U. H. Gerlach, ``Quantum States of a Field Partitioned by an Accelerated Frame'', Phys. Rev. D {\bf 40}, 1037-1047 (1989)}
 
\bib {S. A. Fulling, Phys. Rev. D$\underline {7}$, 2850 (1973).}
\smallskip

\bib {W. Israel, Phys. Lett $\underline {57}$A, 107 (1976).}
\smallskip

\bib {W. G. Unruh, Phys. Rev. D $\underline {14}$, 870 (1976).}
\smallskip

\bib {See, for example, N. J. Vilenkin, $\underline {\rm Special\,\,Functions\,\,
and\,\,the\,\,Theory\,\,of\,\,Group}$ \hfil\break $\underline {\rm
Representations}$, (American Mathematical Society Providence, R. I.,
1968) p.195-198, 248-252.}
\smallskip

\bib { U. H. Gerlach, "Minkowski Bessel Modes", Phys. Rev. D {\bf 38},
514-521 (1988); gr-qc/9910097 . }

\end